# Ensemble learning based linear power flow

Ren Hu
Dept. of Electrical & Computer Engineering
University of Central Florida
Orlando, USA
hurenlaker@knights.ucf.edu

Qifeng Li
Dept. of Electrical & Computer Engineering
University of Central Florida
Orlando, USA
Qifeng.li@ucf.edu

*Abstract*—This paper develops an ensemble learning-based linearization approach for power flow, which differs from the network-parameter based direct current (DC) power flow or other extended versions of linearization. As a novel data-driven linearization through data mining, it firstly applies the polynomial regression (PR) as a basic learner to capture the linear relationships between the bus voltage as the independent variable and the active or reactive power as the dependent variable in rectangular coordinates. Then, gradient boosting (GB) and bagging as ensemble learning methods are introduced to combine all basic learners to boost the model performance. The fitted linear power flow model is also relaxed to compute the optimal power flow (OPF). The simulating results of standard IEEE cases indicate that (1) ensemble learning methods outperform PR and GB works better than bagging; (2) as for solving OPF, the data-driven model excels the DC model and the SDP relaxation in the computational accuracy, and works faster than ACOPF and SDPOPF.

*Index Terms*-- Power flow, data-driven, ensemble learning, gradient boosting, bagging.

## I. INTRODUCTION

Power flow analysis is an indispensable part for power system planning and operation in practice. However, the nonlinear and nonconvex properties of power flow equations restrict the convergency and computational speed, especially in large scale power systems, which attracts many researchers to put forward treatments related to the challenges above. One of the typical measures is linearizing the alternating current (AC) power flow, such as the direct current (DC) power flow that maps the linear relationship between the active power and the voltage phase angle. Other extended linearized models of power flow can also be found in [1]- [7].  [1] proposes in distribution networks linear approximations of the active and reactive power demands, and inferences the sufficient conditions for the existence of the power flow solution. Similarly, in [2], a linear load flow for three-phase power distribution systems is proposed with the consideration of balanced and unbalanced operations as ZIP models of the loads. Other linearized models through the decoupled voltage magnitude and phase angle as independent variables can be further seen in [3]-[7].

Although these approaches enhance the accuracy beyond the DC power flow, most of them are through simplifying network parameters to reach the linearization of power flow, or exploiting variable-selection based methods [8] that indirectly change the bus number, even replace the original voltage variables by new variables and lead to the poor applicability in optimization and control problems. In the presence of massive databases and sufficient measurements related to power system operation, data-driven approaches have gradually attracted more and more attentions, applied in estimating Jacobian matrix [9] distribution factors [10], and the admittance matrix [11]. To this end, from the perspective of data mining, the polynomial regression [12], [13] is exercised as a basic learner to learn the linear associations between the bus voltage as the independent variable and the active or reactive power as the dependent variable. Meanwhile, gradient boosting [14], [15] and bagging [16], [17] as ensemble learning techniques are leveraged to assemble all basic learners to improve the model performance. The learned linear power flow model is also utilized to compute the optimal power flow (OPF), manifesting its better computational accuracy than the DC model and the semi-definite positive (SDP) relaxation method, and its faster computational speed than ACOPF and SDPOPF.

## II. PROBLEM FORMULATION AND STATEMENT

This section depicts the existing problems and solution in computing power flow, as well as some challenges and tactics in fitting power flow from the perspective of data mining.

### A. Existing Problems and the Solution

The nonlinearity and nonconvexity of AC power flow have interested researchers in the topics of simplifying power flow and designing algorithms with better convergency. Though the linearizations of power flow based on the assumptions of simplifying network parameters, to some extent, accelerate the power flow computation [1]- [8], their accuracy cannot match the SDP relaxation. In the consideration of advancing the computational accuracy and efficiency of algorithms and making the full use of big data techniques, the data mining method seems to be worth exploring with the accessible data pertaining to power system operation and computational resources.

### B. Fitting Challenges and Remedies

One of most common problems in machine learning is overfitting models indicated by a good performance on the training dataset yet a bad performance on the test dataset.

This work is supported by the U.S. National Science Foundation under Grant #1808988.

Correspondingly, ensemble learning [17], [18] approaches are recommended to avoid overfitting when learning the linear power flow model, through regularizing learning parameters. Another fitting issue is to handle the multicollinear correlations within the bus voltages as the independent variables, which is actually caused by insufficient data to reveal their true associations. Although many variable selection or shrinkage methods [19] have been suggested to meliorate the multicollinearity through reducing the number of independent variables, they may indirectly change the bus number and remove crucial variables for further optimization and control application in power system. Therefore, increasing more datasets for fitting models can be an appropriate method to prevent the multicollinearity and retain the bus number and key parameters in models without removing variables.

## III. ENSEMBLE LEARNING BASED LINEAR POWER FLOW

This part consists of the mapping rules of linear power flow, ensemble learning methodologies and the formulation of OPF based on the fitting linear models.

### A. Linear Mapping Formulations

Based on the data mining techniques, the nonlinear quadratic forms of AC power flow in a $n$-bus system can be transformed into linear representations represented as

$$P_i = A_i X + b_i$$
$$Q_i = C_i X + d_i$$
$$P_{ij} = A_{ij} X_{ij} + b_{ij} \quad (1)$$
$$Q_{ij} = C_{ij} X_{ij} + d_{ij}$$

where $P_i$, $Q_i$ are the active or reactive power at $i$-th bus; $P_{ij}$, $Q_{ij}$ are the active or reactive power at $ij$-th branch ($i, j = 1,2,...,n$); $A_i$, $C_i$, $A_{ij}$, $C_{ij}$ are the linear coefficient vectors; $b_i$, $d_i$, $b_{ij}$, $d_{ij}$ are the constant terms of linear formulations; $X = [x_1\ x_2,..., x_{2n}]$ is the bus voltage vector formulated in rectangular coordinates; $x_i$ denotes the real or imaginary part of the $i$-th bus voltage.

### B. Ensemble Learning Methodologies

Gradient boosting (GB) and bagging as the typical ensemble learning methods are leveraged in this paper to reinforce the polynomial regression (PR) as a basic learner and compute all linear coefficient vectors and constant terms in (1). Assume that there is a dataset containing $M$ samples $\{(X_m, Y_m)\}_{m=1}^{M}$ where $X_m = [x_{m1}, x_{m2}, .., x_{m(2n)}]$ is the $m$-th sample of bus voltages as the independent variables; $Y_m$ is the $m$-th sample of the active or reactive power at all buses or branches, for instance, $Y_m = [p_{m1}, p_{m2},..., p_{mn}] = \{p_{mi}\}$ is the $m$-th sample of the active power at all buses. We take the dependent variable $P_i$ as a general example to illustrate how to apply GB and bagging.

#### 1. Gradient Boosting

The general work of GB is to tweak learning parameters in an iterative fashion to find the minimum descending gradient through minimize a loss function. We choose the mean squared error function as the specific loss function in (2)

$$L(p_{mi}, \widehat{p_{mi}}) = \frac{1}{2}(p_{mi} - \widehat{p_{mi}})^2 \quad (2)$$

where $p_{mi}$ and $\widehat{p_{mi}} = p_{mi}(X)$ are the observed and estimated values of $P_i$. The procedure of GB is depicted as below.

---

**Algorithm: gradient boosting**

1. Initialize the model with a constant value $p_{mi}^0(X)$.

$$p_{mi}^0(X) = arg\ \min_{\alpha} \sum_{m=1}^{M} L(p_{mi}, \alpha) \quad (3)$$

where $\alpha$ is the initial constant vector.

2. For $t = 1$ to $T$ where $T$ is the number of learners.
   1) Compute the descending gradient $\gamma_t$ by

$$\gamma_t = -\left[\frac{\partial L(p_{mi}, p_{mi}(X))}{\partial p_{mi}(X)}\right]_{p_{mi}(X) = p_{mi}^{t-1}(X)} \quad (4)$$

   2) Fit a base learner $\varphi_t(X; \rho)$ by

$$\rho_t = arg\ \min_{\theta} \sum_{m=1}^{M} L(\gamma_t, \varphi_t(X_m; \rho)) \quad (5)$$

   where $\rho_t$ is the coefficient vector of $\varphi_t(X; \rho)$ by fitting $\gamma_t$. Here the polynomial regression (PR) as a basic learner is used to fit the key parameters in equations (1).

   3) Compute the learning rate $\theta$ by

$$\theta_t = arg\ \min_{\theta} \sum_{m=1}^{M} L(p_{mi}, p_{mi}^{t-1}(X_m) + \theta \varphi_t(X; \rho)) \quad (6)$$

   In practice, we can also set a constant learning rate. The smaller $\theta$ is applied, the better generalization is achieved.

   4) Update the model:

$$p_{mi}^t(X) = p_{mi}^{t-1}(X) + \theta_t \varphi_t(X) \quad (7)$$

3. Output $p_{mi}^T(X)$

---

#### 2. Bagging

Bagging as a special case of the model averaging approach can reduce variances and avoid overfitting by tweaking the number of bootstraps. The key of bagging is to draw random samples with replacement and combine a basic learning method to train models. The algorithm is written as below.

---

**Algorithm: bagging**

1. For $bt = 1$ to $BT$ where $BT$ is the number of bootstraps.
   1) At the $bt$-th bootstrap draw $M'$ ($M' \leq M$) random samples with replacement.
   2) Fit a base learner $p_{mi}^{bt}(X; \rho)$ by

$$\rho_{bt} = arg\ \min_{\rho} \sum_{m=1}^{M} L(p_{mi}^{bt}, p_{mi}^{bt}(X_m; \rho)) \quad (8)$$

   where $p_{mi}^{bt}$ denotes the observed value of $P_i$ at the $bt$-th bootstrap; $\rho_{bt}$ is the coefficient vector of $p_{mi}^{bt}(X; \rho)$ by fitting $p_{mi}^{bt}$. Similarly, the PR as the basic learner is performed to compute all parameters in equations (1).

2. Output $p_{mi}^{bag}(X)$ by averaging all bootstrap results in

$$p_{mi}^{bag}(X) = \frac{1}{BT} \sum_{bt=1}^{BT} p_{mi}^{bt}(X) \quad (9)$$

where $p_{mi}^{bag}(X)$ is the predictive value of $P_i$.

---

### C. Convexifying the Optimal Power Flow

According to the fitted linear power flows, the data driven convex relaxation (DDCR) for OPF can be rewritten in (10) where $G$ is the index set of generators; $c_{i0}$, $c_{i1}$, $c_{i2}$ are the $i$-th generator cost coefficients; $P_i^G$, $Q_i^G$ are the $i$-th generator active and reactive power; $P_i^L$, $Q_i^L$ are the active and reactive power load at $i$-th bus; $P_i^{Gmin}$, $P_i^{Gmax}$, $Q_i^{Gmin}$, $Q_i^{Gmax}$ are the lower and upper limits of the $i$-th generator active and reactive power; $V_i^{max}$, $S_{ij}^{Gmax}$ are the maximums of the $i$-th bus voltage and the $ij$-th branch transmission capacity; $x_{2i-1}$ and $x_{2i}$ ($e_i$ and $f_i$) are the real and imaginary part of voltage.

Minimize $\sum_{i \in G} (c_{i0} + c_{i1} P_i^G + c_{i2} P_i^{G2})$

$$s.t. \begin{cases} A_i X + b_i \leq P_i^G - P_i^L \\ C_i X + d_i \leq Q_i^G - Q_i^L \\ x_{2i-1}^2 + x_{2i}^2 \leq V_i^{max2} \\ P_i^{Gmin} \leq P_i^G \leq P_i^{Gmax} \\ Q_i^{Gmin} \leq Q_i^G \leq Q_i^{Gmax} \\ P_{ij}^2 + Q_{ij}^2 \leq S_{ij}^{Gmax} \\ A_{ij} X_{ij} + b_{ij} \leq P_{ij} \\ C_{ij} X_{ij} + d_{ij} \leq Q_{ij} \end{cases} \quad (10)$$

## IV. SIMULATION ANALYSIS

In this study, Monte Carlo method is exercised to generate random data samples of the bus voltage, the active and reactive power at each bus or branch, given in IEEE 5, 57 and 118-bus systems [20]. The active and reactive power loads are stochastically fluctuating around 0.6~1.1 times of preset values. Generally, the empirical required minimum sample size is at least 2.4 times as many as the number of buses [8], [10], [19].

### A. Comparing Performance

The simulation outcomes through the polynomial regression (PR), gradient boosting (GB), and bagging, are obtained by the equal size of training and test datasets. Here the average root mean square error (RMSE) of the predicted dependent variable is defined to symbol the predictive accuracy, and the performance demonstration of all methods is characterized by comparing the test RMSEs, not the training RMSEs, shown in TABLE I.

TABLE I TEST AND TRAINING RMSES OF ALL METHODS ON DIFFERENT CASES

| Case | Method | PR | | GB | | Bagging | |
|---|---|---|---|---|---|---|---|
| RMSE (10 e-05) | | test | training | test | training | test | training |
| case 5 (size=175, T=200, BT=50) | $P$ | 9476.04 | 11.22 | 64.08 | 59.01 | 310.04 | 112.91 |
| | $Q$ | 2247.02 | 556.30 | 395.06 | 353.26 | 1333.24 | 570.53 |
| | $P_{ij}$ | 248.11 | 76.50 | 43.92 | 39.81 | 144.66 | 88.62 |
| | $Q_{ij}$ | 1812.00 | 323.00 | 179.32 | 175.83 | 385.72 | 341.62 |
| case 57 (size=250, T=200, BT=50) | $P$ | 67.30 | 4.63 | 17.27 | 6.5 | 26.43 | 5.90 |
| | $Q$ | 237.99 | 18.14 | 59.36 | 22.55 | 88.93 | 19.77 |
| | $P_{ij}$ | 23.52 | 17.20 | 17.27 | 6.51 | 19.73 | 18.04 |
| | $Q_{ij}$ | 63.04 | 42.50 | 52.65 | 52.39 | 54.51 | 50.09 |
| case 118 (size=400, T=200, BT=50) | $P$ | 100.33 | 15.87 | 41.65 | 17.12 | 82.17 | 16.07 |
| | $Q$ | 180.27 | 30.45 | 79.06 | 32.22 | 161.17 | 31.31 |
| | $P_{ij}$ | 56.97 | 20.80 | 20.64 | 20.71 | 25.90 | 21.56 |
| | $Q_{ij}$ | 117.57 | 57.09 | 63.24 | 62.56 | 75.44 | 58.60 |

Note that the unit of above data is 10 e-05.

From TABLE I, some observations can be observed on all cases: 1) ensemble learning methods work better than the single PR; 2) GB outperforms bagging and PR; 3) there is no pronounced overfitting as the training RMSE is consistently smaller than the test RMSE for any dependent variable.

### B. Tuning Parameters

As the model performance described by the test and training RMSEs is directly associated with a set of regularization parameters to handle overfitting, this part presents the tuning processes of GB and bagging, respectively, with regard to the number of learners T and the number of bootstraps BT.

#### 1. Tuning the Number of Learner T in GB

The model performances of the active and reactive power injections $P$ and $Q$ on all cases are plotted in the logarithms of RMSEs in Fig. 1. (a)~(c). Each plot depicts how the RMSEs change as the number of learners T increases. From Fig. 1, we can observe that:

(1) for all cases, the test and training RMSEs of $P$ and $Q$ gradually decrease to stable levels with the increase of the number of learner T;

(2) each case reaches the balance points at different numbers of learner. For case 5, after T=150 and T=180 the test and training RMSEs of $Q$ and $P$ tend to be constant, separately. The RMSEs of case 57 become hardly changeable when T=100 for $Q$ and T=150 for $P$. Similarly, the RMSEs of case 118 stop descending when T=140 for $Q$ and T=180 for $P$;

(3) No evident overfitting or underfitting problems are observed on all cases through the tuning process.

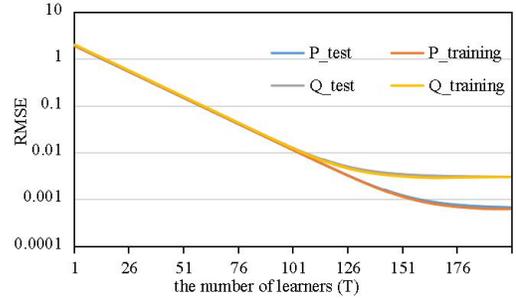
(a) case 5: RMSEs of $P$, $Q$ by boosting(log)

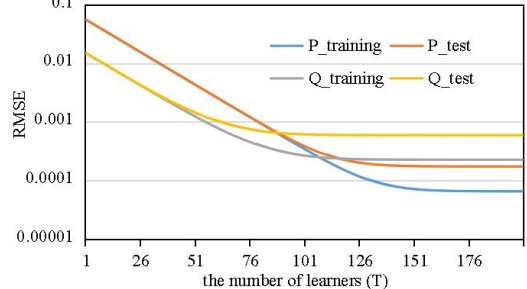
(b) case 57: RMSEs of $P$, $Q$ by boosting(log)

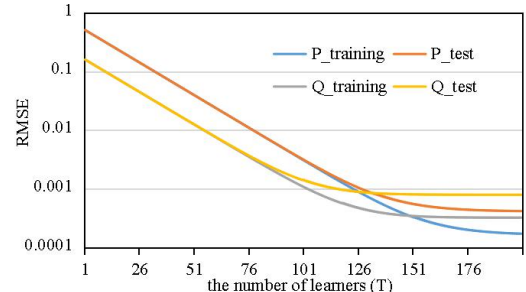
(c) case 118: RMSEs of $P$, $Q$ by boosting(log)

Fig. 1. The test and training RMSEs of $P$, $Q$ with increasing T

#### 2. Tuning the Number of Bootstraps BT in Bagging

In Fig. 2. (a)~(f), comparing the RMSEs of the active power injection $P$ on all cases before and after bagging is incorporated. Each plot represents the results of each bootstrap (blue broke curve) and bagging (red curve) as the number of bootstrap BT goes up. According to Fig. 2, we can inference that:

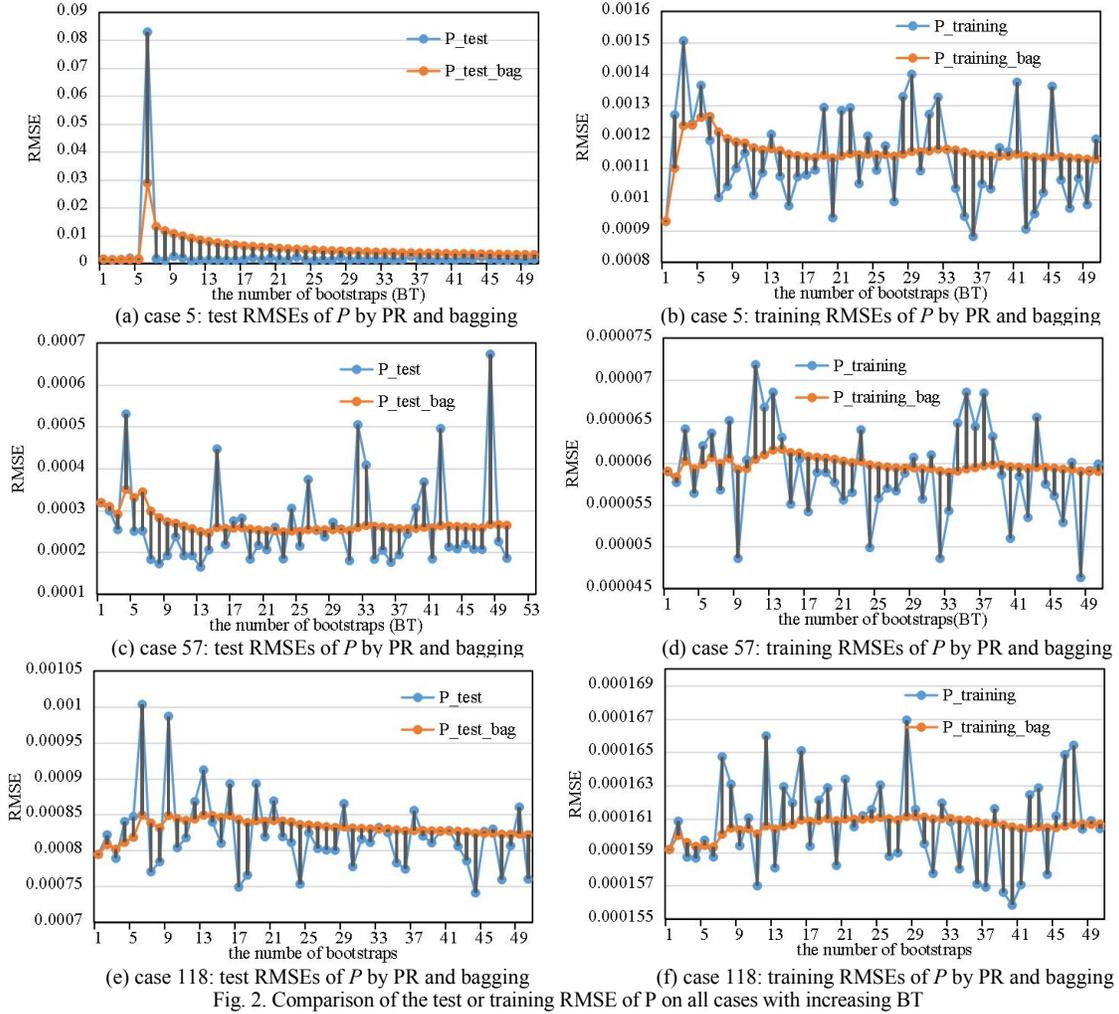

Fig. 2. Comparison of the test or training RMSE of P on all cases with increasing BT

(1) the test or training RMSE of bagging tends to stablize with slight fluctuations after sufficient times of bootstraps. When the number of bootstrps is 20 or more, both the test and training RMSEs seem to work well on all cases.

(2) the outcome of every single bootstrap fluctuates stochastically around the red curve, indicating that the single learner has its unstable weakness.

(3) Bagging can average the variances of all single learners and avoid overfitting.

## C. Comparing OPF Results

As GB exhibits its superiority over others, the linear models fitted by GB are chosen to develop a data-driven convex relaxation (DDCR) method and compute the optimal objective values (OOV) shown in TABLE II (unit: $/hr), compared with the original nonconvex ACOPF, the DCOPF and the semidefinite programming relaxation of OPF (SDPOPF). Note that the results of ACOPF are set as the benchmarks and the optimality gap is defined as: (the difference with ACOPF)/ACOPF denoted in TABLE III. Additionally, the runtime (unit: s) of each approach is recorded in TABLE IV.

TABLE II COMPARING OBJECTIVE VALUES OF OPF

| case | ACOPF | DDCROPF | DCOPF | SDPOPF |
|---|---|---|---|---|
| case 5 | 17551.89 | 17547.4 | 17479.9 | 16635.78 |
| case 57 | 12100.86 | 12096.04 | 10211.99 | 10458.06 |
| case 118 | 129660.70 | 129680.13 | 125947.88 | 129713.07 |

TABLE III COMPARING OPTIMALITY GAPS

| case | ACOPF | DDCROPF | DCOPF | SDPOPF |
|---|---|---|---|---|
| case 5 | 0% | 0.025% | 0.41% | 5.22% |
| case 57 | 0% | 0.040% | 15.61% | 13.57% |
| case 118 | 0% | -0.014% | 2.86% | -0.040% |

TABLE IV COMPARING RUNTIMES OF DIFFERENT METHODS

| case | ACOPF | DDCROPF | DCOPF | SDPOPF |
|---|---|---|---|---|
| case 5 | 2.95 | 1.57 | 2.3 | 23.82 |
| case 57 | 2.83 | 2.15 | 1.77 | 31.99 |
| case 118 | 3.94 | 2.91 | 2.11 | 39.08 |

From TABLE II~ IV, they reveal that:

(1) For the accuracy of OOV, DDCR performs better than the DC method and the SDP relaxation on all cases.

(2) The table of optimality gaps proves that DDCR (-0.014%~0.040%) works more robustly than the SDP relaxation (-0.040%~13.57%) and the DC method (0.41%~15.61%) on the computational accuracy.

(3) TABLE IV indicates on all cases DDCROPF runs more efficiently than ACOPF and SDPOPF, and its runtimes even can match DCOPF.

## V. CONCLUSIONS AND FUTURE WORK

In this paper, an ensemble learning based linearization of power flow is proposed, unlike the conventional DC model and other extended linear versions according to the approximations of simplifying network parameters or variable selection methods that remove some independent variables. The proposed approach applies the polynomial regression as a basic leaner to characterize the linear associations between the bus voltages as the independent variables and the active or reactive power at buses and branches as the dependent variables, without removing the key independent variables. Gradient boosting and bagging as ensemble learning methods are exploited to combine all basic learners to boost the model performance. Eventually, the fitted linear power flow models are relaxed to compute the optimal power flow and compared with the conventional DC method, the SDP relaxation and the nonconvex AC power flow. The experimental outcomes reveal that ensemble learning methods indeed excel the single learner and gradient boosting works the best; in terms of computing the optimal power flow, the fitted data driven model outperforms the DC method and the SDP relaxation on the computational accuracy, and runs faster than ACOPF and SDPOPF.